# Pronounced photovoltaic response from multi-layered transition-metal dichalcogenides PN-junctions

Shahriar Memaran,[§,†] Nihar R. Pradhan,[§] Zhengguang Lu, [§,†] Daniel Rhodes,[§,†] J. Ludwig,[§,†] Q. Zhou,[§,†] Omotola Ogunsolu,[ห] Pulickel M. Ajayan,[x] Dmitry Smirnov,[§] Antonio I. Fernández-Domínguez, [#] Francisco J. García-Vidal, [#] Luis Balicas,[§,*]

[§]National High Magnetic Field Lab, Florida State University, 1800 E. Paul Dirac Dr. Tallahassee, FL 32310, USA.
[†]Department of Physics, Florida State University, Tallahassee, Florida 32306, USA.
[ห]Department of Chemistry & Biochemistry, Florida State University Tallahassee, FL 32306-4390 USA.
[x]Department of Mechanical Engineering and Materials Science, Rice University, Houston, TX 77005-1892, USA.
[#]Departamento de Física Teórica de la Materia Condensada and Condensed Matter Physics Center (IFIMAC), Universidad Autónoma de Madrid, E-28049 Madrid, Spain.

ABSTRACT: Transition metal dichalcogenides (TMDs) are layered semiconductors with indirect band gaps comparable to Si. These compounds can be grown in large area, while their gap(s) 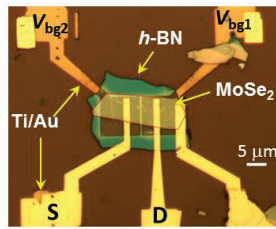 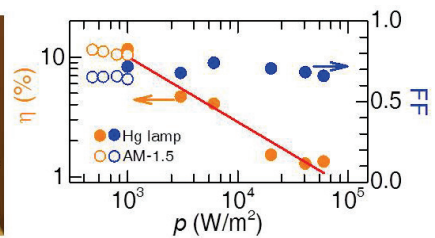 can be tuned by changing their chemical composition or by applying a gate voltage. The experimental evidence collected so far, points towards a strong interaction with light, which contrasts with the small photovoltaic efficiencies $\eta \leq 1\%$ extracted from bulk crystals or exfoliated monolayers. Here, we evaluate the potential of these compounds by studying the photovoltaic response of electrostatically generated PN-junctions composed of approximately ten atomic-layers of $MoSe_2$ stacked onto the dielectric $h$-BN. In addition to ideal diode-like response, we find that these junctions can yield, under AM-1.5 illumination, photovoltaic efficiencies $\eta$ exceeding 14 %, with fill-factors of ~ 70 %. Given the available strategies for increasing $\eta$ such as gap tuning, improving the quality of the electrical contacts, or the fabrication of tandem cells, our study suggests a remarkable potential for photovoltaic applications based on TMDs.

**KEYWORDS**: *transition metal dichalcogenides, molybdenum diselenide, PN-junctions, photovoltaic-effect, fill-factor.*

The photovoltaic-effect (PE), or the creation of a voltage, or an electrical current, in a given material or a solution upon light exposure, was discovered by Becquerel[1] in 1839. Nevertheless, the effective energy harvest from sunlight only became possible in the 1950's with the advent of the silicon PN-junctions.[2] PN junctions are adjacent hole- and electron-doped semiconducting regions having an interface depleted of charge-carriers. PN-junctions are fundamental building blocks for today's electronics and optoelectronics whose fundamental technology is still based on Si despite recent progress in, for example, perovskite solar cells.[3-5]

Several of the transition metal dichalcogenides (TMDs) such as $MoS_2$, $WSe_2$, etc., are semiconducting, but van der Waals bonded solids which are exfoliable down to a single atomic layer.[6,7] These compounds, and their heterostructures, can be grown in high quality and in wafer size area.[8,9] Monolayers display unique optical[10-12] as well as optoelectronic properties[13,14] owing to their direct band gap.[10] It was recently shown[15-20] that it is possible to observe current rectification and the photovoltaic- effect in photodiodes electrostatically built through two lateral back-gates to form a horizontal or lateral PN-junctions from a monolayer,[17-19] or from heterostructures composed of atomically thin TMDs in combination with graphene.[15,16,20,21] In contrast to Si, thin layers of TMDs are inherently flexible, semi-transparent, and lack interfacial dangling bonds which, as argued in Ref.[20] would allow the creation of high-quality heterointerfaces without the constraint of atomically precise commensurability. The availability of TMDs with distinct band gaps[7,22] and work functions, opens the possibility of i) engineering the band gaps in heterostructures, and ii) fabricating translucent photovoltaic tandem cells composed of TMDs having distinct gaps therefore absorbing photons with energies ranging from the ultra-violet (e.g. $HfS_2$) to the infrared (e.g. $WSe_2$, $MoSe_2$, or $MoTe_2$).

For *horizontal* PN junctions based on a single atomic layer of $WSe_2$ under white light illumination, Ref.[17] reports maximum short-circuit current densities $j_{sc}$ ~ 0.23 A/cm$^2$ (defined as the photo-generated current in *absence* of a bias voltage flowing from the junction towards the electrical contacts) for a an illumination power density $p$ = 1.4 x 10$^3$ W/m$^2$. Despite this anomalously large $j_{sc}$ value, this junction yields a quite modest photovoltaic efficiency η = 0.5 %. In turn, Ref.[13] reports $j_{sc}$ ≅ 150 A/cm$^2$ under a fairly large power density $p$ ≅ 3.2 x 10$^6$ W/m$^2$ yielding a number of photo-generated carriers circulating through the photodetector per adsorbed photon and per unit time, or external quantum efficiency (EQE), of just ~10$^{-3}$. Small EQEs would suggest rather small power conversion efficiencies.

For *vertical* heterojunctions composed of *single* atomic layers of $MoS_2$ (*n*-doped) and $WSe_2$ (either *p*-doped or ambipolar), Ref.[20] reports $j_{sc}$ values approaching 1 mA/cm$^2$ under $p$ = 10$^6$ W/m$^2$ (laser light with λ = 532 nm). Ref.[21] on the other hand, reports $j_{sc}$ ≅ 13 mA/cm$^2$ under white light illumination but with a concomitant small η of 0.2 %. For multi-layered stacks of $MoS_2$ and $WSe_2$ contacted with graphene Ref.[20] reports an incredibly high $j_{sc}$ value of ~ 2.2 A/cm$^2$ acquire under laser light (λ = 532 nm) and very high power densities. even larger $j_{sc}$ values were reported for heterostructures composed of graphene acting as electrodes, and multilayered $MoS_2$,[16] under laser light with extremely large $p$s (>10$^6$ W/m$^2$). These observations suggest that junctions composed of multi-layered TMDs can yield higher photovoltaic currents than monolayers, leading perhaps to higher power conversion efficiencies. Thicker crystals would allow longer photon travel distances within the material thus increasing the probability of generating electron-hole pair(s).

To evaluate this hypothesis, we fabricated lateral PN junctions (see Fig. 1, as well as Methods) based on exfoliated *h*-BN (with thicknesses *t* ranging between ~20 and ~40 nm) on top of which we transferred[23] chemical vapor transport synthesized $MoSe_2$ single-crystals, previously found by us to display ambipolar behavior[24]. Although, a similar architecture was already reported,[17-19] here we i) evaluate the properties of a different compound, i.e. MoSe$_2$ and ii) evaluate the potential of TMDs for photovoltaic applications by focusing on bulkier crystals. We find that our photodiode is found to exhibit photovoltaic power conversion efficiencies approaching 14 % under standard AM-1.5 solar spectrum. This value is not far from those extracted from the best Si solar cells,[25] i.e. 25%, and compare favourably with those of transparent photovoltaic cells.[26-28] This efficiency is likely to increase, for example, by varying the band gap when tuning the composition, by changing the materials used for the contacts[15,16], the incorporation or plasmonic nanoparticles,[29] or the optimization of the number of atomic layers. Given their relative transparency in the visible region,[30] and the ability to grow large areas,[8,9,31] we argue that few layer transition metal dichalcogenides present a remarkable potential for photovoltaic applications.

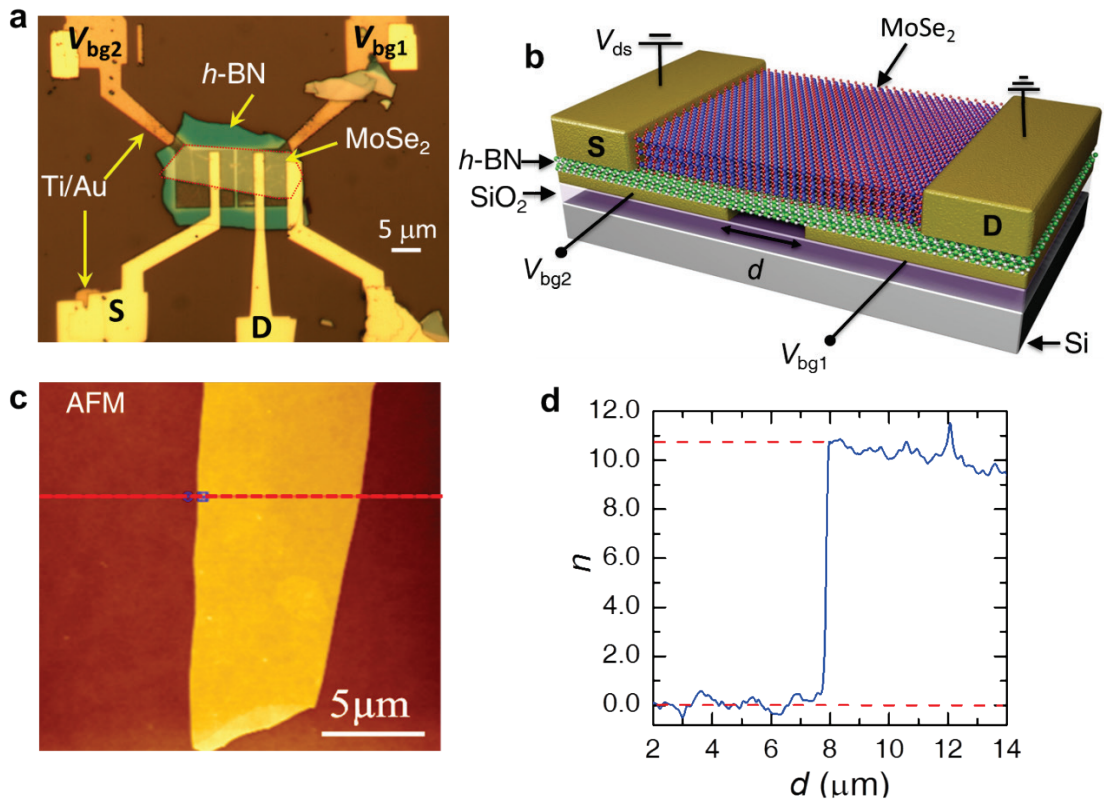

**Figure 1. a** Micrograph of a multi-layered MoSe$_2$ crystal stacked onto a ~ 45 nm thick *h*-BN crystal (sample #1), itself placed on a double gate structure patterned on a SiO$_2$/*p*-Si substrate. A thin red dotted line is used to delineate the MoSe$_2$ crystal. The channel length (in between both voltage leads

and at either side of the junction) is $l$ =5.52 µm and its average width $w$ = 6.85 µm. The gap between gates is $\cong$ 400 nm. **b** Sketch of a typical sample, indicating the configuration of measurements, e.g. drain (D) and source (S) contacts, excitation voltage $V_{ds}$, as well as the voltages applied to the back-gates, $V_{bg1}$ and $V_{bg2}$, respectively. Contacts and gates are composed of 50 nm of Au deposited onto a 5 nm thick layer of Ti. **c** Atomic force microscopy (AFM) image of the MoSe$_2$ crystal on SiO$_2$. Red line depicts the line along which the height profile shown in **d** was collected. **d** Height profile for the mechanically exfoliated MoSe$_2$ crystal showing the number of layers $n = h/c'$, where $h$ is the height of the flake and $c' = c/2$ = 6.4655 Å is the inter-layer separation ($c$ is the lattice constant along the inter-planar direction[32]).

Figure 1a shows a photomicrograph of one of our MoSe$_2$ crystals stacked on $h$-BN in a lateral PN junction configuration. Micrographs of additional samples measured for this study are shown in the Supporting Information. As seen, the MoSe$_2$ crystal is perceptible, but is transparent enough to allow the visualization of both back gates through the $h$-BN crystal whose thickness was determined to be ~ 45 nm through atomic force microscopy (AFM). Figure 1b shows a schematic of the measurements, where three independent voltages are applied to the sample, respectively $V_{bg1}$ for the left back-gate, $V_{bg2}$ for the right one, and the bias voltage $V_{ds}$ through the source and drain contacts. One measures the resulting drain to source current $I_{ds}$ under or without illumination. Figure 1c shows an AFM image of the MoSe$_2$ crystal. The red line indicates the line along which the height profile shown in Fig. 1d was collected, which divided by an inter-layer spacing[32] of 6.4655 Å indicates a crystal composed of 10 to 11 atomic layers.

Figure 2a displays the absolute value of the drain to source current $|I_{ds}|$ for a second multi-layered MoSe$_2$ or sample # 2 (8.8 µm wide and ~13 atomic layers thick) as a function of the excitation voltage $V_{ds}$, when 6 V is applied to either gates but with opposite polarity. As expected for a PN junction, and as previously reported for single-layer WSe$_2$ heterostructures,[17-19] our multi-layered MoSe$_2$ heterostructure displays rectification, or a diode-like response with the sense of current rectification being dependent upon the gate-voltage profile across the junction. Red lines are fits of the observed diode response to the Shockley equation in the presence of a series resistor:[33]

$$I_{ds} = \frac{fV_T}{R_s} W_0\left(\frac{I_0 R_s}{fV_T} \exp\left(\frac{V + I_0 R_s}{fV_T}\right)\right) - I_0 \quad (1)$$

where $W$ is the Lambert function and $V_T$ the thermal voltage, yielding an ideality factor $f$ of 1.4 with values for the series resistance $R_s$ ranging from 0.25 to ~0.5 MΩ. These $f$ values are smaller than those in Refs.[17,18] ($1.9 \leq f \leq 2.6$) for single layered WSe$_2$ lateral diodes suggesting a diode response that is closer to the ideal one. The Shockley-Read-Hall recombination theory,[34,35] which assumes recombination via isolated point defect levels, predicts $f \leq 2$.

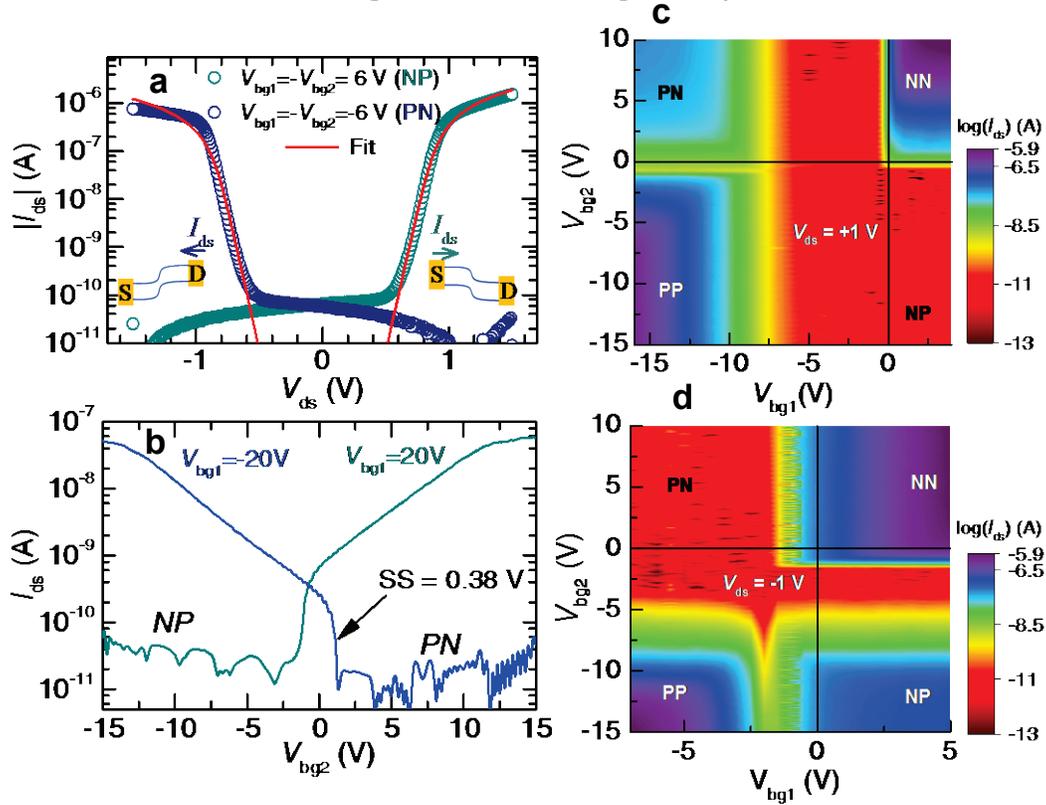

**Figure 2. a** Absolute value of the drain to source current |$I_{ds}$| on a log-log scale and as a function of the excitation voltage $V_{ds}$ for two configurations of the back-gate voltages, i.e. $V_{bg1} = - V_{bg2} = 6$ V which corresponds to the accumulation of electrons (N) and holes (P) at the respective interfaces, or NP configuration (dark cyan markers), and -$V_{bg1} = V_{bg2} = 6$ V or PN configuration (blue markers), respectively. Notice the diode-like response as a function of either positive or negative values of $V_{ds}$, depending on the sign of the gate voltage(s). Notice the factor of $10^5$-$10^6$ increase in current. Red lines are fits to the Shockley diode equation including a series resistance $R_s$. For the $V_{ds} < 0$ V branch one extracts an ideality factor $f$ = 1.4 with $R_s$ = 0.45 MΩ. For the $V_{ds} > 0$ V we obtain $f$ = 1.4 with $R_s$ = 0.28 MΩ. This data was collected on sample #2 (13 atomic layers). **b** Field-effect characteristics (from sample #1) obtained by keeping the excitation voltage $V_{ds}$, and one of the gate-voltages $V_{bg1}$ at constant values of 0.3 and 20 V respectively, and by sweeping the second gate-voltage $V_{bg2}$. As clearly seen, the field-effect response is

ambipolar, i.e. one can accumulate either electrons (for positive values for *both* gate-voltages) or holes (e.g. negative values for both gate-voltages) in the channel. **c** Contour plot of the logarithm $|I_{ds}|$ as a function of both gate voltages and for an excitation voltage $V_{ds}$ = +1 V. This data was collected from sample #1. Notice the clear ambipolar response when both gate voltages have the same polarity or, the rectification-like response when they have opposite polarities. **d** Same as in **c** but for $V_{ds}$ = -1 V. By comparing **c** and **d**, one observes a clear asymmetry in the NP response with respect to the PN one, due to the gate-voltage induced diode response.

Figure 2b displays $I_{ds}$ as a function of the gate voltage $V_{bg2}$ while $V_{bg1}$ is maintained at a fixed value of +20 V (dark cyan trace) and -20 V (blue trace), respectively. As seen, a sizeable drain to source current is observed only when both gate voltages have the same polarity, due to the electric-field induced accumulation of charge carriers (transistor operation). The so-called sub-threshold swing, or SS ~ 380 mV per decade, is considerably sharper than those values previously observed by us[24] for MoSe$_2$ on SiO$_2$. The same observation applies to the threshold gate voltage for conduction, i.e. the voltage beyond which one extracts a sizeable current, which is nearly one order of magnitude smaller for *h*-BN substrates, or between 1 and 2 V. Not surprisingly, both observations indicate that *h*-BN is a superior substrate, i.e. less disordered, when compared to SiO$_2$.

Finally, Figs. 2c and 2d display contour plots of the logarithm of the drain to source current $I_{ds}$ as a function of both gate voltages, and for excitation voltages $V_{ds}$ = +1 V and -1 V, respectively. This data set was collected from the sample #1. Currents ranging between $10^{-12}$ and $10^{-11}$ A, which correspond to our noise floor, are depicted by the red regions in both plots. Currents ranging from ~ 0.1 to 1 µA are depicted by the clear- and darker-blue regions, respectively. It is clear that a sizeable current is obtained when *both* gates are simultaneously energized with the *same* voltage due to the field-effect induced accumulation of charges in the channel. However, both figures become asymmetric when the gates are energized with opposite polarities: sizeable currents are observed in the second and in fourth quadrants of Figs. 2c and 2d, respectively. As expected for PN-junctions, this indicates current rectification but whose diode response is controllable by the relative polarity between both gate voltages. Having established a well-defined diode response, we proceed with the characterization of their photovoltaic response.

As seen in the Supplementary Figure S3, we evaluated the *intrinsic* photovoltaic power conversion efficienciy of our MoSe$_2$ crystals (i.e. in absence of back gate-voltage(s)) and under laser illumination ($\lambda$=532 nm, spot diameter $\cong$ 3.5 μm) finding that it is remarkably small ~$10^{-3}$ %. As explained in the Supplementary Information this contrasts markedly with the power conversion efficiencies obtained when both gates are energized to generate the PN-junction. Supplementary Figures S4 and S5 display a thorough characterization of our PN-junctions under laser illumination. We observe very high power conversion efficiencies, i.e. $\eta = P^{el}_{max}/P_i \approx 40$ %, for incident illumination power densities approaching $p_i$ = 1000 W/m$^2$, where $P^{el}_{max}$ corresponds to the maximum photo-generated electrical power. Here, to calculate the incident power $P_i$, and similarly to Ref. 17, we multiplied $p_i$ by the *active* area of the junction, or $A_j = w_c$ x $w_j$, where $w_c$ is the width of the crystal and $w_j$ is the gap between both back-gates or the region between gates which is depleted from charge carriers. Although, based on Fig. 1a, one could argue that the channel and related PN-junction, might extend well beyond the gap between gates, through nearly the entire channel. Hence, the active area $A_j$ might end up being considerably larger than the values calculated by us, and thus necessarily yielding smaller $\eta$ values. Here, and as shown below, we address this issue by producing PN-junctions whose top metallic contacts cover nearly the entire channel area although still allowing us to illuminate the area of the depleted junction. In addition, we will focus on the evaluation of the photovoltaic power conversion efficiency under the standard Air Mass 1.5 (AM1.5) spectrum, which is meaningful from a technological perspective than laser illumination.

Figure 3a shows an optical micrograph of a FET whose channel area is covered by wide drain and source electrical contacts, with the intention of extracting the intrinsic photovoltaic response of the depleted junction.

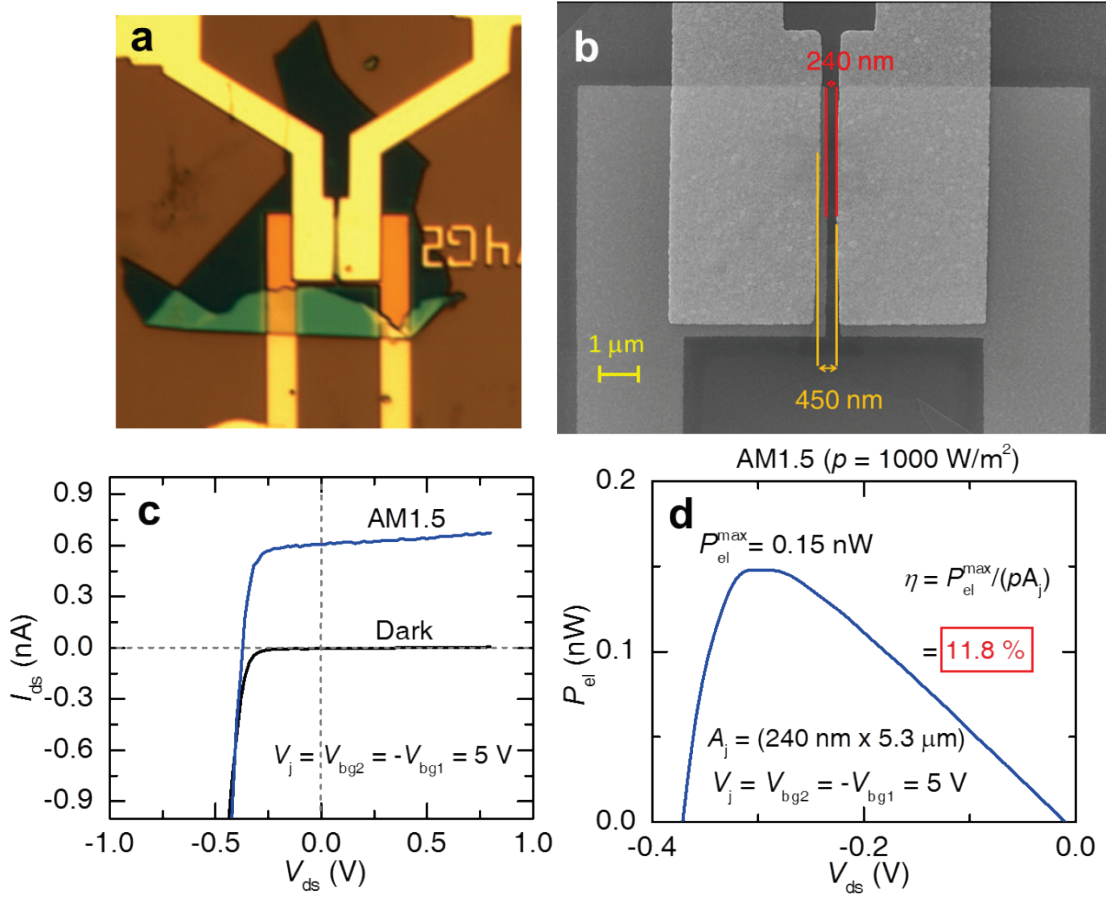

**Figure 3. a** Optical micrograph of sample #3. **b** Scanning electron microscopy image indicating the width $w_j$ = 240 nm of the depletion junction and the separation $l$ = 450 nm between drain and source contacts. **c** Drain to source current $I_{ds}$ as a function of the bias voltage $V_{ds}$, when $V_{bg}$ = + 5V is applied to one of the gates and $V_{bg}$ = − 5 V is applied to the other. The response in absence of illumination, or under dark conditions, is depicted by the black line while the response under AM 1.5 spectrum is depicted by the blue line. **d** Photogenerated power $P_{el}$ = $I_{ds}$ x $V_{ds}$ as a function of $V_{ds}$ yielding a maximum value of 0.15 nW. When normalized by the illumination power shone onto the junction, i.e. 1000 W/m$^2$ x (240 x 10$^{-9}$ x 5.3 x 10$^{-6}$) m$^2$, one obtains a power conversion efficiency η = 11.8 %. If instead, one used the channel length $l$ = 450 nm, one would obtain η = 6.3 %.

Given the limited precision of our e-beam lithography system, in Fig. 3b we show a scanning electron microscopy image of the same FET, indicating both the final separation $w_j$ = 240 nm between the back gates and the average separation, i.e. $d$ = 450 nm, between both top electrodes. Figure 3c shows a comparison between the diode-response extracted from this sample, both under AM1.5 (blue line) and in absence of illumination (black trace), when each gate is energized under a constant value of 5 V but of opposite polarity. As seen, under AM1.5 illumination power density (1000 W/m$^2$), one extracts a short-circuit current $I_{sc}$ = 606 pA (0.1143 nA/µm) which contrasts markedly with the values obtained for $V_{ds}$ > 0 V under dark conditions (i.e. oscillating between 1 and 5 pA, and corresponding to our noise floor). Figure 3d on the other hand, shows the photogenerated electrical power $P_{el}$ = $I_{ds}$ x $V_{ds}$ as a function of $V_{ds}$. This curve is obtained by subtracting the black trace from the blue one, which subtracts from the calculated electrical power the contribution from the source-meter. As seen, it peaks at a maximum value $P_{el}^{max}$ = 0.15 nW which yields a conversion efficiency η ≅ 11.8 %, when renormalized by the illumination power $P_i$ = $p_i$ x $A_j$ = 1000 W/m$^2$ x ($A_j$ = 240 nm x $w_c$ = 5.3 µm) shining on the surface $A_j$ of the depleted junction. If instead, one used the power applied to the entire channel one would obtain η ≅ 6.3 %. This value is 12.6 times larger than η = 0.5 % reported for a lateral PN-junction based on a WSe$_2$ monolayer,[17] also larger than the value η = 5.3 % reported for MoS$_2$ monolayers composing a type-II heterojunction with p-Si,[36] or larger than η = 2.8 % obtained from plasma doped[37] MoS$_2$. For this sample one obtains an open circuit voltage $V_{oc}$ = $V(I_{ds} = 0\ A)$ = 0.364 V yielding a fill factor FF = $P_{el}^{max}$/($I_{sc}$ x $V_{oc}$) = 0.68 which is comparable to the values obtained for conventional Si based solar cells.

Metallic surface plasmon polaritons are known to dramatically enhance the interaction between carriers and light in small structures such as sample#3, which have a separation between metallic electrodes comparable to, or smaller than the characteristic wavelength of light.[29,38] Surface plasmon polaritons (SPPs) can propagate back and forth between the metal terminations, effectively creating a Fabry–Perot resonator.[38] This effect could lead to considerably higher photovoltaic power conversion efficiencies[29] thus contributing to the high efficiencies observed by us. In order to quantify the role of SSPs and their contribution to the above quoted efficiency, we have performed numerical simulations of the light transmission process, to evaluate how much of the incident light is transmitted to the active area defined by the MoSe$_2$ layer, which is associated with the photovoltaic response of our devices. By using COMSOL Multiphysics, a commercially available solver of Maxwell equations, we were able to calculate the transmission spectra for both *p* (along the channel length) and *s* (polarization perpendicular to the channel length) polarized light, evaluated at the different interfaces of our samples. The details concerning our numerical calculations are presented in the Supplementary Information. The main outcome of these simulations is that only about 70 % of the incident light is impinging at the gap between both back gates acting on the MoSe$_2$ depleted junction. In other words, for the geometry of our sample, we find that SPPs would be detrimental for its photovoltaic conversion efficiency. Our simulations imply that the actual illumination power density irradiated onto the depleted area is ~ 720 W/m$^2$ leading to a power conversion efficiency $\eta \cong 16.4\%$.

In order to confirm the correct value of $\eta$ we also evaluated samples having geometries which are similar to that of sample #1 (see Fig. 1a) but with shorter channel lengths. By varying the length of the channel one should be able to determine if the effective area of the junction extends beyond the depletion area between both gates or, if the only relevant factor when evaluating $\eta$ is truly the depleted area. Here, we evaluated two samples, having similar crystal widths of $w_c = 7.5$ μm but distinct thicknesses; sample # 4 (12 atomic layers) and characterized under the white light spectrum produced by a Hg lamp and sample # 5 (8 atomic layers) characterized under AM1.5.

In Fig. 4a we show an optical image of sample # 5 while Fig. 4b shows its scanning electron microscopy image, from which we extract the precise dimensions of its depleted area, i.e. $w_j$ = 313 nm, and a crystal width $w_c$ = 7.5 µm. Both samples, were previously characterized under coherent $\lambda$ = 532 nm laser light, yielding very similar $\eta$ values with respect to the ones extracted from samples #1 and #2. Again, to evaluate $\eta$ we used the dimensions of the carrier depleted area and not the length of the channel which would yield dissimilar $\eta$ values. This will be illustrated below through a comparison between photovoltaic efficiencies extracted under AM1.5 from samples #3 and #5.

Figure 4c presents the photo-diode response observed from sample #4 under the spectrum of an Hg lamp namely, $I_{ds}$ as a function of $V_{ds}$ when the back gates are energized under $V_j = V_{bg} = \pm$ 4 V. With the Hg lamp one can precisely vary $p_i$ from 0 to 60 kW/m$^2$. Under $p_i$ = 1000 W/m$^2$, one extracts $I_{sc}$ = 836 pA or 0.1115 nA/µm which is just ~ 2.5% smaller than the short circuit current extracted from sample #3 under AM1.5 and $V_{bg} = \pm$ 5 V. This indicates that an increase in the separation between the electrical contacts by a factor > 6 did not lead to any substantial decrease in the photo-generated electrical current due to electron-hole recombination. Figure 4d displays the photo-generated electrical power $P_{el}$ extracted from the traces in Fig. 4c and as a function of $V_{ds}$. As before, to evaluate $P_{el}$ ($V_{ds}$, $p$) we subtracted $P_{el}$ ($V_{ds}$, $p$ = 0.0 W/m$^2$), which should eliminate any spurious contribution from the power supply.

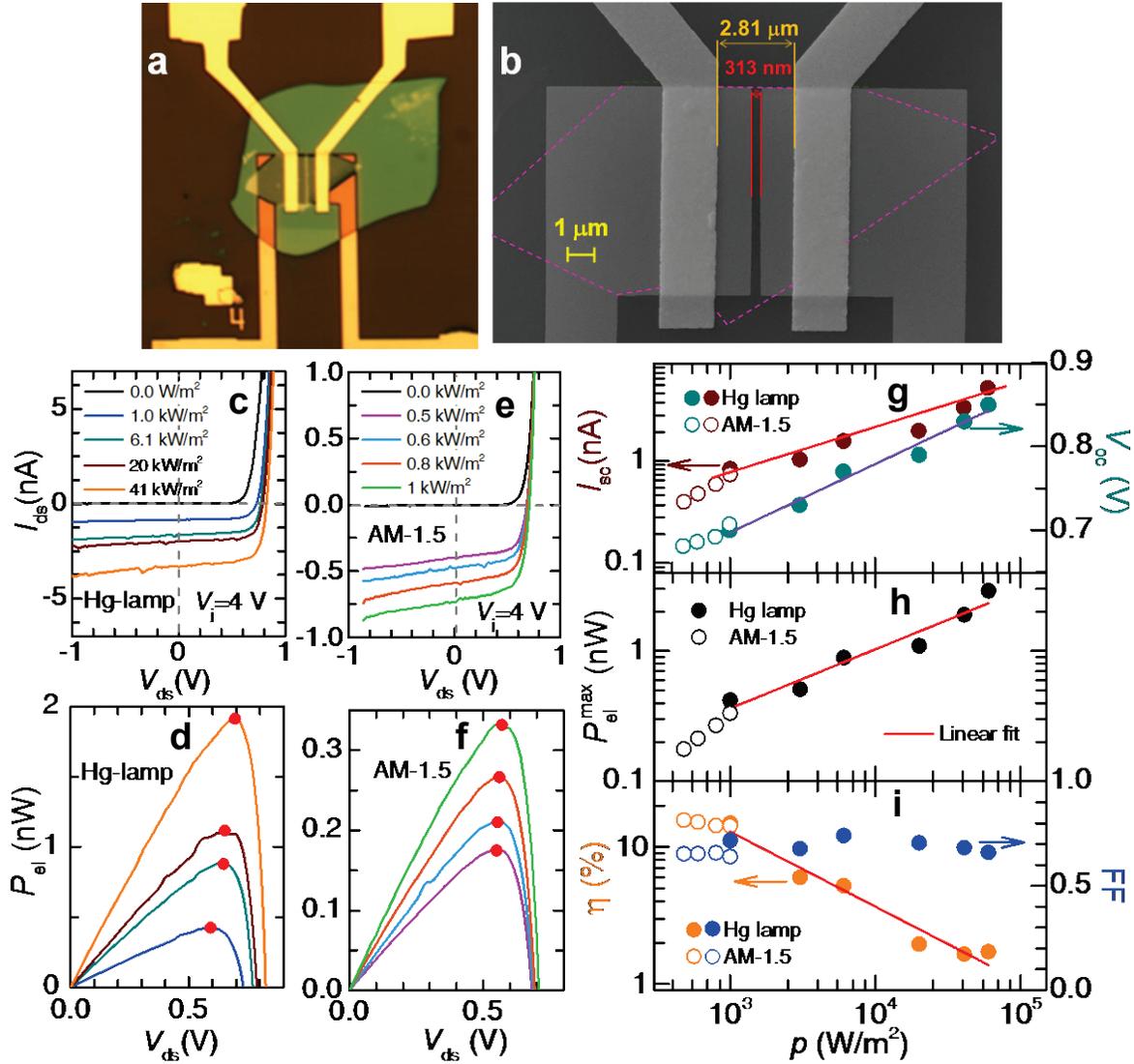

**Figure 4. a** Optical micrograph of sample #5 which is characterized by a larger separation $l$ between contacts. **b** Scanning electron microscopy image indicating the width of the depletion junction or $w_j$ = 313 nm, as well as $l$ = 2.81 μm. **c** Drain to source current $I_{ds}$ as a function of bias voltage $V_{ds}$ under several illumination power densities $p$, under the spectrum of a Hg-lamp, and for a multilayered MoSe$_2$ based PN-junction (sample #4). **d** From the curves in **c** concomitant photo-generated electrical power $P_{el}$ = $I_{ds}$ x $V_{ds}$ as a function of $V_{ds}$. Here, and for each curve, $P_{el}$ was calculated by subtracting the $p$ = 0.0 W/m$^2$ data (black line). Red markers indicate the maximum photo-generated electrical power values $P_{el}^{max}$. **e** $I_{ds}$ as a function of $V_{ds}$ under AM-1.5 spectrum and also for lower $p$ values using the same set-up. **f** $P_{el}$ as a function of $V_{ds}$, from the curves in **c** and calculated in a similar manner. **g** Log-log plot of the short circuit current $I_{sc}$ (brown markers) and semi-log plot of the open circuit voltage $V_{oc}$ (cyan makers) as functions of $p$. Red line is a linear fit of $I_{ds}(p)$ while the violet line corresponds to a semi-logarithmic fit of $V_{oc}(p)$. **h** $P_{el}^{max}$ as a function of $p$, from the red markers in **d** and **f**. Red line is a linear fit. **i** Photovoltaic efficiency η (orange markers) and fill factor FF (blue markers) as functions of $p$. For panels **g**, **h** and **i** solid and open markers indicate values measured under an Hg lamp and under AM-1.5 irradiation, respectively.

Under $p_i$ = 1000 W/m$^2$ one obtains $P_{el}^{max}$ = 0.42 nW which once renormalized by $P_i$ = 1000 W/m$^2$ x ($A_j$ = 7.5μm x 400 nm) = 2.3475 nW yields η = 14.01 %. Notice that the typical values for the leakage current flowing through the gates is < 10$^{-11}$ A. Hence, when multiplied by $V_j$ = 4 V one would obtain $P_{bg}$ ≅ 0.04 nW (<< $P_{el}^{max}$) which would correspond to the *maximum* power applied to the back-gates. The evaluation of the photovoltaic response of sample #5 under AM1.5 is presented in Figs. 4e and 4f which present $I_{ds}$ and $P_{el}$ as functions of $V_{ds}$, respectively. Under AM 1.5 ($p_i$ = 1000 W/m$^2$), one extracts $I_{sc}$ = 738 pW (or 0.098 nA/μm) and a $P_{el}^{max}$ = 0.334 nW. When using $A_j$ to calculate $P_i$ one obtains η = 14.23 % which is very close to η = 14.01 % obtained from sample #4 and to η = 11.8 % (or 16.4 % as the simulations imply) extracted from sample #3. Hence all three samples yield consistent η values under white light illumination.

Instead, if one used the area of the channel (between the electrodes) to calculate the $P_i$ illuminating sample #5, one would obtain η = 1.58%. This value is 4 times smaller than η = 6.3 % calculated for sample #3 in a similar way. Given that i) their short circuit currents (in nA/μm) differ by only ~15% which implies a similar density of photo-generated electron-hole pairs, ii) that $P_{el}^{max}$ for sample #5 is 2.2 times larger than the value extracted from sample #3, iii) and that sample #3 exposes a much smaller area of the active material, it would be unphysical to obtain a 4 times higher power conversion efficiency for this sample. Therefore, we conclude that the correct calculation of η ought to be based on the active area of the carrier depleted junction which yields *consistent* values ranging between ~14 and ~ 16 %. In any case, we have solidly established η = 6.3 % as the bare minimum power conversion efficiency under AM1.5 for a PN-junction composed solely of a transition metal dichalcogenide as the semiconducting channel material. Although our simulations imply that the actual value is $\eta \cong 8.75$ %.

Figure 4g displays $I_{sc}$ (brown markers) and the extracted open circuit voltages $V_{oc}$ (dark green markers) values extracted under the white light spectrum produced by a Hg lamp (solid markers) and under the AM1.5 spectrum (open symbols). Notice that in either case, under the standard power density of ~ $10^3$ W/m², one would obtain a short circuit current density of $j_{sc}$ ~ 1 A/cm² if one normalized $I_{sc}$ by the cross-sectional area of the MoSe$_2$ crystal. As for any solar cell, this is the current flowing from the junction towards the electrical contacts. However, our geometry is distinct, with the light laterally impacting the junction. Conventional solar cells are vertical stacks of *n*-doped and *p*-doped material whose top surface is exposed to light. Hence, our $j_{sc}$ values cannot be directly compared with those extracted from conventional solar cells, which oscillate around 40 mA/cm². Nevertheless, we hope that our observations will stimulate theoretical efforts addressing the significance of such pronounced $j_{sc}$ values.

$V_{oc}$ on the other hand is observed to range from ~ 0.7 to ~0.85 V which is comparable to $V_{oc}$ ~ 0.7 V which is a typical value[25] for high quality monocrystalline Si solar cells. The red line is a power law fit, i.e. $I_{sc} \propto p^{\gamma}$ yielding $\gamma \cong 0.4$ while the brown line is a logarithmic fit of $V_{oc}(p)$. Figure 4h displays $P_{el}^{max}$ as a function of *p*, as obtained from the traces in both Figs. 4d (solid markers) and 4f (open markers) with the red line being a power law fit, yielding again an exponent of $\cong 0.4$. Finally, Fig. 4g displays the resulting $\eta = P_{el}^{max}/P_i$ and the concomitant fill factor FF, where $p_i$ was multiplied by the active area of the depletion junction in order to calculate $P_i$. $\eta$ displays power-law dependence as a function of *p*. More importantly, under the standard AM-1.5 spectra ($p$ ~ $10^3$ W/m²), $\eta$ would surpass ~14 % which is ~24 times larger than the maximum value observed for monolayers[17] and exceeds by one order of magnitude values previously reported for bulk transition metal dichalcogenides.[39] As previously discussed, and as shown in the Supplementary Information, these $\eta$ values cannot be attributed to the intrinsic photovoltaic response of MoSe$_2$ or to the exposed areas adjacent to the PN junction, since our studies indicate that these areas would yield negligible or very small contributions to the photovoltaic efficiencies reported here.

In summary, our studies on the *intrinsic* photovoltaic response of multilayered MoSe$_2$ field-effect transistors yield photovoltaic power conversion efficiencies $\eta$ well below 1%. These values are in general agreement with previous reports based on bulk[39] and on transition metal dichalcogenides single-atomic layers.[17,40] Nevertheless, when a MoSe$_2$ crystal composed of ~10 atomic layers is transferred onto a flat *h*-BN crystal, itself placed on a pair of lateral back-gates to create an electrostatic PN-junction, one observes photovoltaic efficiencies surpassing 14% under AM-1.5 spectrum, with concomitant fill factors approaching 0.7. These *non-optimized* values compare well with those of current Si technologies and with organic tandem solar cells.[41]

An important aspect requiring immediate theoretical attention is to understand the anomalously large short current densities > 1 A/cm$^2$ extracted from the lateral geometry used here, which surpass by far those observed of conventional vertically stacked solar cells.[25]

Finally, the sharp increase in efficiency relative to single atomic layers[17,19] is attributable to the increase in sample thickness. This implies that a systematic study as a function of the number of atomic layers is required to expose the maximum photovoltaic efficiencies extractable from these materials. The current challenge is to translate these efficiencies onto large area, vertically stacked heterostructures. Notice that an indirect gap of 1.41 eV[42] for multi-layered MoSe$_2$ would yield a maximum photovoltaic efficiency of $\eta \sim 35$ % (for a single PN junction) according to the Shockley–Queisser limit[43]. While tandem cells composed of transition metal dichalcogenides having distinct band gaps would not be subjected to this limit. Coupled to our results, this implies a remarkable potential for the use of transition metal dichalcogenides in photovoltaic applications specially it these required flexibility and light transmittance.[30]

ASSOCIATED CONTENT

*Supporting Information available:* It contains the methods section, describing sample synthesis, device fabrication, and experimental set-up examples of fits to the Schokley diode equation with a series resistance, micrographs of samples #2, and characterization of the intrinsic photovoltaic response of a multi-layered $MoSe_2$ crystal under laser light ($\lambda$ = 532 nm), and the results of our simulation which evaluates the role of surface plasmon polaritons on the transmission of light towards the depleted junction. This material is available free of charge *via* the Internet at http://pubs.acs.org.

AUTHOR INFORMATION

Corresponding Author
  Email: balicas@magnet.fsu.edu

**Author Contributions**: LB conceived the project. DR synthesized and characterized the single crystals. SM produced the devices. NP and SM performed electrical transport characterization under illumination with the help of ZL, JL, and DS. Measurements under AM1.5 were performed with the support of OO. AIFD and FJGV performed the numerical simulations in order to evaluate geometry of our samples and potential role of surface plasmon polaritons. The manuscript was written through contributions of all authors. All authors have given approval to the final version of the manuscript.

**Notes**: The authors declare no competing financial interest.

*Acknowledgement.* We acknowledge Prof. P. Kim and Dr. C.-H. Lee for their guidance on the development of our crystal transfer and stacking technique(s), Prof. K. Hanson for allowing access to the Solar Simulator and Dr. K. Emery, for the critical reading of this manuscript. This work was supported by the U.S. Army Research Office MURI Grant No. W911NF-11-1-0362. J. L. acknowledges the support by NHMFL UCGP No. 5087. Z.L .and D.S. acknowledge the support by DOE BES Division under grant no. DE-FG02-07ER46451. F.J.G-V. acknowledges support from the European Research Council (ERC-2011-AdG proposal no. 290981). The NHMFL is supported by NSF through NSF-DMR -1157490 and the State of Florida.

# Supporting Information for manuscript titled "Pronounced photovoltaic response from multi-layered transition-metal dichalcogenides PN-junctions"

Shahriar Memaran,[§,†] Nihar R. Pradhan,[§] Zhengguang Lu,[§,†] Daniel Rhodes,[§,†] J. Ludwig,[§,†] Q. Zhou,[§,†] Omotola Ogunsolu,[ʜ] Pulickel M. Ajayan,[ˣ] Dmitry Smirnov,[§] Antonio I. Fernández-Domínguez,[#] Francisco J. García-Vidal,[#] Luis Balicas,[§,*]

[§]National High Magnetic Field Lab, Florida State University, 1800 E. Paul Dirac Dr. Tallahassee, FL 32310, USA.
[†]Department of Physics, Florida State University, Tallahassee, Florida 32306, USA.
[ʜ]Department of Chemistry & Biochemistry, Florida State University Tallahassee, FL 32306-4390 USA.
[ˣ]Department of Mechanical Engineering and Materials Science, Rice University, Houston, TX 77005-1892, USA.
[#]Departamento de Física Teórica de la Materia Condensada and Condensed Matter Physics Center (IFIMAC), Universidad Autónoma de Madrid, E-28049 Madrid, Spain.
E-mail:balicas@magnet.fsu.edu

## 1. Methods

i. *Sample synthesis:* $MoSe_2$ single crystals were synthesized through a chemical vapor transport technique using either iodine or excess Se as the transport agent. 99.999% pure Mo powder and 99.999% pure Se pellets were introduced into a quartz tube together with 99.999% pure iodine. The quartz tube was vacuumed, brought to 1150 $^o$C, and held at this temperature for 1.5 weeks at a temperature gradient of < 100 $^o$C. Subsequently, it was cooled to 1050 $^o$C at a rate of 10 $^o$C per hour, followed by another cool down to 800 $^o$C at a rate of 2 $^o$C per hour. It was held at 800 $^o$C for 2 days and subsequently quenched in air.

ii. *Device fabrication:* Multi-layered flakes of $MoSe_2$ were exfoliated from these single crystals by using the micromechanical cleavage technique, and transferred onto *p*-doped Si wafers covered with a 270 nm thick layer of $SiO_2$. *h*-BN crystals (Momentive PolarTherm PT110) were mechanically exfoliated from larger crystals, and transferred onto pre-evaporated Ti:Au gates, using a technique similar to the one described in Ref.(*22*). The $MoSe_2$ crystal was subsequently transferred onto the *h*-BN crystal by using the same technique. For making the gates and the electrical contacts 90 nm of Au was deposited onto a 4 nm thick layer of Ti *via* e-beam evaporation. Contacts and gates were patterned by using standard e-beam lithography techniques. After each transfer, as well as after the final gold deposition, the samples are annealed at 200 $^0$C for ~ 2 h in forming gas. After the heterostructure is completed it is re-annealed under high vacuum for 24 hours at 120$^o$C.

iii. *Measurements and experimental set-up:* Atomic force microscopy (AFM) imaging was performed using the Asylum Research MFP-3D AFM. Electrical characterization was performed by using a sourcemeter (Keithley 2612 A) coupled to a Physical

Property Measurement System. For short- or photo-current measurements a Coherent Sapphire 532-150 CW CDRH and Thorlabs DLS146-101S were used, with a continuous wavelength of 532 nm. Light was transmitted to the sample through a 3 μm single-mode optical-fiber with a mode field diameter of 3.5 μm. The size of the laser spot was also measured against a fine grid. Hence, here we use 3.5 μm for the laser spot diameter assuming a constant power density distribution in order to approximate the Gaussian distribution corresponding to the mode field diameter of 3.5 μm (see Supplementary Information). For the white light measurements, samples were irradiated with an AM 1.5 solar spectrum generated from a 300 W Xe arc lamp (Ushio, UXL-302-O) enclosed in a Oriel Research Arc Lamp Housing (Newport, 67005) with the light output passed through a AM 1.5 Global filter (Newport, 81094) and mechanical shutter (Newport, 71445). The light intensity was measured using a Calibrated reference cell and meter (Newport, 91150V) and the intensity was adjusted by using neutral density filters. An Hg lamp was used for white light illumination power densities beyond 1000 W/m$^2$. The incident illumination power was controlled by using neutral UV-VIS filters placed between the lamp and the sample. An aperture was used to define a spot diameter of ~9.3 mm. A broadband OPHIR-3A detector was used to measure the illumination power density.

2. Fit to the Schokley diode equation with series resistance

In Figure S1 below we show a couple of examples of fits of the observed diode response to the Shockley equation in the presence of a series resistor $R_s$, which as mentioned in the main text is particularly sensitivity on the value of the reverse bias current $I_0$. An $I_0$ value of $10^{-12}$ A, as observed experimentally, yields a poor fit as seen in the right panel of the supplementary figure S1, where markers depict experimental data and red line the actual fits for both diode rectification branches. It also yields remarkably large values for the diode ideality factor $f > 3.4$, although with reasonable values for the shunt resistance $R_s$, i.e. between 0.5 and 1.7 MΩ. We find that good fits are obtained when $I_0$ is allowed to decrease to values approaching at least $10^{-15}$ A as shown in the right panel of Fig. S1. In this case, the ideality factor approaches a value of 2 with Rs between 0.8 and 2.4 MΩ.

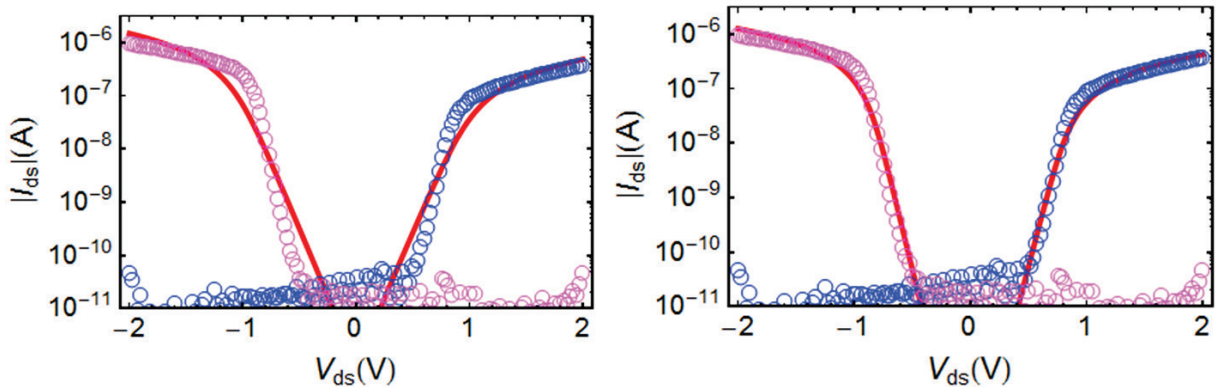

**Figure S1.** Right panel: Logarithm of the extracted drain to source current $I_{ds}$ as a function of the excitation voltage $V_{ds}$ for both NP (magenta markers) and PN (blue markers) configurations of the back gates. Red lines are fits to Schokley diode equation with series resistor for a value $I_0 \sim 10^{-12}$ A. One obtains diode ideality factors $f = 3.5$ for PN, (with $R_s = 1.7$ MΩ) and $f = 3.4$ for NP (with $R_s = 0.5$ MΩ). Left panel: same as in the right panel but for $I_0 \sim 10^{-15}$ A. The fit yields $f = 2.1$ for PN ($R_s = 2.4$ MΩ), and $f = 1.9$ for NP ($R_s = 0.8$ MΩ).

3. **Micrograph of Sample #2**

In Figure S2 below we show the picture of sample #2 in the main manuscript, whose MoSe$_2$ crystal is composed of approximately 13 atomic layers according to an AFM height profile, with an average width of 8.86 µm. Separation between contacts is ~ 3.35 µm. The thickness of the h-BN crystal is approximately ~ 30 nm.

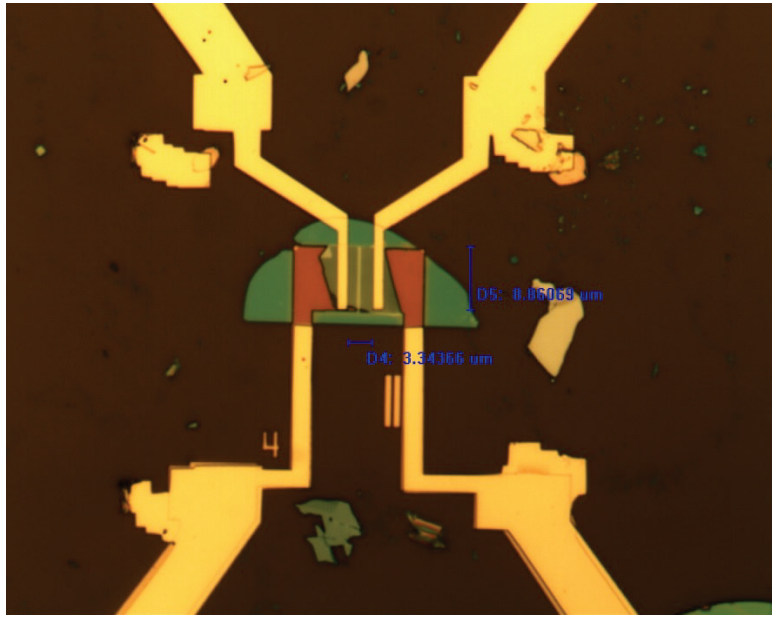

**Figure S2**. Micrograph of a second 13 layers thick MoSe$_2$ crystal on $h$-BN.

4. **Characterization of a MoSe$_2$ on h-BN heterostructure under laser illumination and in absence of gate voltage: photovoltaic response.**

In Figure S3 we evaluate the photo-response of multi-layered MoSe$_2$ on $h$-BN in absence of gate voltages. Figure S3a displays the generated photocurrent $I_{ph} = I_{ds}(P) - I_{ds}(P = 0 \text{ W})$ as a function of the excitation voltage $V_{ds}$ for several values of the applied laser power $P$ with a spot size = 3.5 µm and for λ = 532 nm. Notice i) the sizeable photo-generated current and ii) the emergence of a finite short-circuit current ($I_{ds}(V_{ds} = 0 \text{ V})$) resulting from the photovoltaic-effect. This indicates i) that the photo-generated electron-hole pairs have a long enough diffusion length to reach the electrical contacts and ii) the existence of a built-in electric-field which drives the photo-generated carriers towards the contacts. This electric-field might result from asymmetric Schottky barriers at the contacts which pins the Fermi-level at distinct positions with respect to the valence band maximum at each contact, thus creating a gradient of chemical potential. This would explain both the photovoltaic response and the asymmetric $I_{ds}$ as a function of $V_{ds}$ characteristics shown in Fig. S3a. Figures S3b and S3c show respectively, the photoresponsivity $R = I_{ph}/P$ and concomitant external quantum efficiency EQE $= hcI_{ph}/e\lambda P$, where $h$ is the Planck constant, $c$ the speed of light, and $e$ the electronic charge, from the data in Fig. S3a.

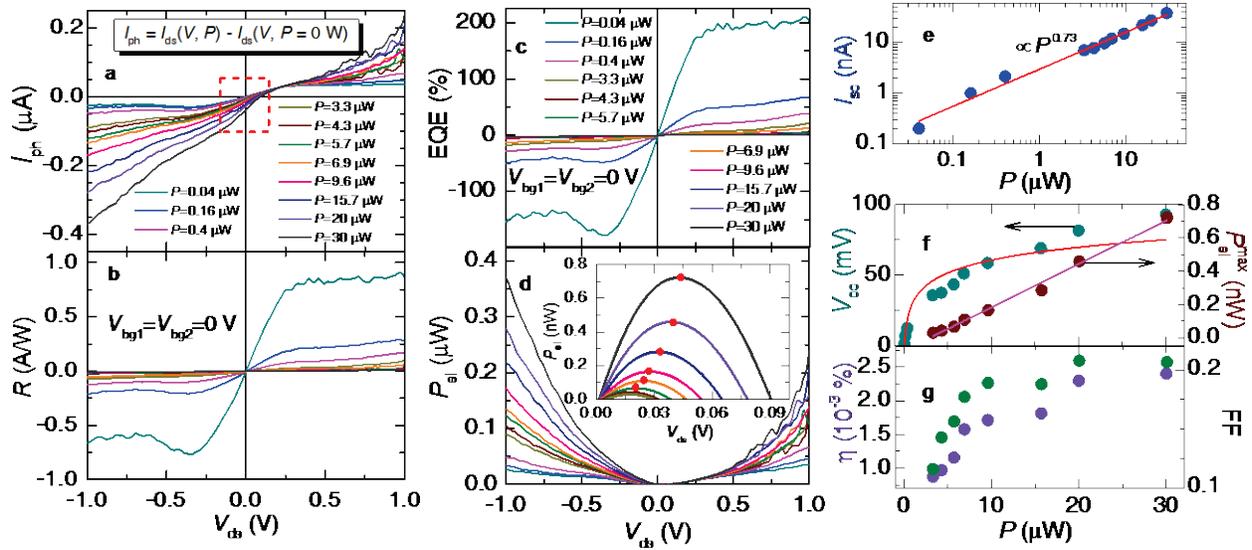

**Figure S3. a** Photocurrent $I_{ph} = I_{ds}$ (P) - $I_{ds}$ ($P = 0$), where $P$ is the applied laser power (wavelength $\lambda = 532$ nm, spot size ~ 3.5 μm, see Methods), as a function of the excitation voltage $V_{ds}$, for several values of $P$ ranging from $P = 0.04$ to 30 μW and in the absence of gate voltage(s). Red square emphasizes the existence of a sizeable, power-dependent, photo-generated current $I_{pv}$ for $V_{ds} = 0$ V, or photovoltaic current. **b** From the curves in (**A**), photoresponsivity $R = I_{ph}/P$ as a function of $V_{ds}$. Notice the quite large values of $R$, approaching 1 A/W at the lower power $P$ levels, but it decreases fast to values ranging from 100 mA/W to 250 mA/W as the power increases by one order of magnitude. **c** External quantum efficiencies from the $I_{ph}$ data in **b**, where $h$, $c$, $e$ and $\lambda$ are the Planck constant, speed of light, electronic charge and excitation wavelength, respectively. Notice that EQE ranges from 25 to ~ 200% at low power levels. **d** Electrical power $P_{el} = I_{ph} \times V_{ds}$ resulting from photon-generated electron-hole pairs as a function of $V_{ds}$ and for several laser $P$ values. As seen, in absence of any gate voltage, a laser power $P = 30$ μW yields ~0.25 and ~0.36 μW of photo generated electrical power for $V_{ds} = -1$ and $+1$ V, respectively. The color code is the same for all four panels: dark cyan trace corresponds to $P = 0.04$ μW, blue trace to $P=0.16$ μW, and so forth. Inset: photogenerated power in the vicinity of zero excitation voltage (in an amplified scale), where it is dominated by the photovoltaic effect. The peak corresponds to the maximum electrical power $P_{el}^{max}$ (red markers) generated by the photovoltaic effect. **e** Short circuit current $I_{sc}$, extracted from **a** when $V_{ds} = 0$ V, as a function of the laser power $P$. Red line is a linear fit yielding $I_{sc} \propto P^{1.3}$. **f** Open circuit voltage $V_{oc}$ (dark cyan markers), as extracted from **a** from the condition $I_{ph} = 0$ A and maximum electrical power $P_{el}^{max}$ from the inset in **d**, both as functions of $P$. Red line is a logarithmic fit. **g** Photovoltaic efficiency $\eta = P_{el}^{max}/P$ (green markers) and fill factor FF $= P_{el}^{max}/ I_{sc}V_{oc}$ (violet markers).

As seen, in absence of gate voltages our MoSe$_2$ on $h$-BN heterostructure(s) shows large photoresponsivities, approaching 1 A/W and EQEs approaching or exceeding 100 %. These values exceed those of heterostructures composed of graphene and multi-layered transition metal dichalcogenides.[1] Figure S3d displays the photo-generated electrical power $P_{el}= I_{ph} \times V_{ds}$ as a function of $V_{ds}$ for several $P$ values. At the lowest $P$ values $P_{ph}$ decreases from > 10 % of $P$ to values approaching just 1 % of $P$ as $P$ increases. The inset shows the electrical power $P_{el}$ resulting solely from the photovoltaic effect (fourth quadrant in a, and for values close to $V_{ds} = 0$ V) where the red dots indicate its maximum values $P_{el}^{max}$. Figure S3e displays the logarithm of the short circuit current $I_{sc}$ (for $V_{ds} = 0$ V) as a function of the logarithm of $P$. Red line is a linear fit from which we extract $I_{sc} \propto P^{\gamma}$ with $\gamma = 0.73$ (1 is the value expected for the photo-thermoelectric effect). Figure S3F displays the extracted open circuit voltage $V_{oc}$ (dark cyan markers) as well as $P_{el}^{max}$ (brown markers) as functions of $P$. Red line is a fit of $V_{oc}$ to a simple logarithmic dependence; for a conventional solar cell $V_{oc} = fk_BT/q \ \ln(I_{ph}/I_0 + 1)$, where $I_0$ is the saturation current under dark conditions, $f$ is the ideality-factor, and $q$ is the electronic charge, while $I_{ph} \propto P$ at low $P$ values. Magenta line is a linear fit of $P_{el}^{max}$ as a function of $P$. Finally, Fig. 3g shows the extracted photovoltaic efficiency $\eta = P_{el}^{max}/P$ (violet markers) and the photovoltaic fill factor FF = $P_{el}^{max}/(I_{sc} \times V_{oc})$ (green markers) as functions of $P$. As seen, $\eta$ is rather small between $1 \times 10^{-3}$ and $2.5 \times 10^{-3}$ %, while FF saturates at FF $\cong 0.2$. Therefore, few layer MoSe$_2$ on $h$-BN, when contacted with Ti:Au, is a low-efficiency photovoltaic architecture (in absence of any gate voltages). Nevertheless, the observation of a photovoltaic response can only be understood in terms of light and gate-induced spatial separation of photo-generated electrons and holes, which increases the exciton recombination times, and of a gradient of the chemical potential due to Fermi level pinning (at the contacts) at distinct positions relative to the bottom of the conduction band.

5. **Photovoltaic response under laser light ($\lambda$ = 532 nm) illumination**

Supplementary Figure S4 describes the overall photovoltaic response of our multi-layered MoSe$_2$ PN-junctions when both gate electrodes are energized and the channel is illuminated with a $\lambda$ = 532 nm laser (with a spot size $\phi \cong 3.5$ μm). Figure S4a plots the short circuit current $I_{sc}$ (or $I_{ds}$ ($V_{ds} = 0$ V)) from sample #1 as a function of $V_j = V_{bg1} = -V_{bg2}$, hence in current rectification mode, for several values of the incident illumination power $P$. As seen, and although the response is asymmetric with respect to $V_j$, one can extract very sizeable short circuit currents (e.g. for $V_j = -5V$) ranging from 0.2 μA to 1.5 μA depending on the $P$ level. This translates into photo responsivities $R=I_{sc}/P$ ranging from 40 to 100 mA/W with concomitant maximum EQEs = $hcR/e\lambda$ ranging from ~10 to 30 % as shown in Fig. S4b.

Figure S4c displays $I_{ds}$ as function of $V_{ds}$, under $V_j = -5$ V for several $P$ levels. Notice the sizeable photo-generated current at zero bias (or the $I_{sc}$) particularly at high $P$ levels. Figure S4d, displays the photo generated electrical power $P_{el} = I_{ds} \times V_{ds}$ due solely to the photovoltaic effect

(i.e. second quadrant in c), where red markers indicate its maximum values $P_{el}^{max}$. These values were obtained after subtraction of the $P = 0.0$ µW data (black trace in c).

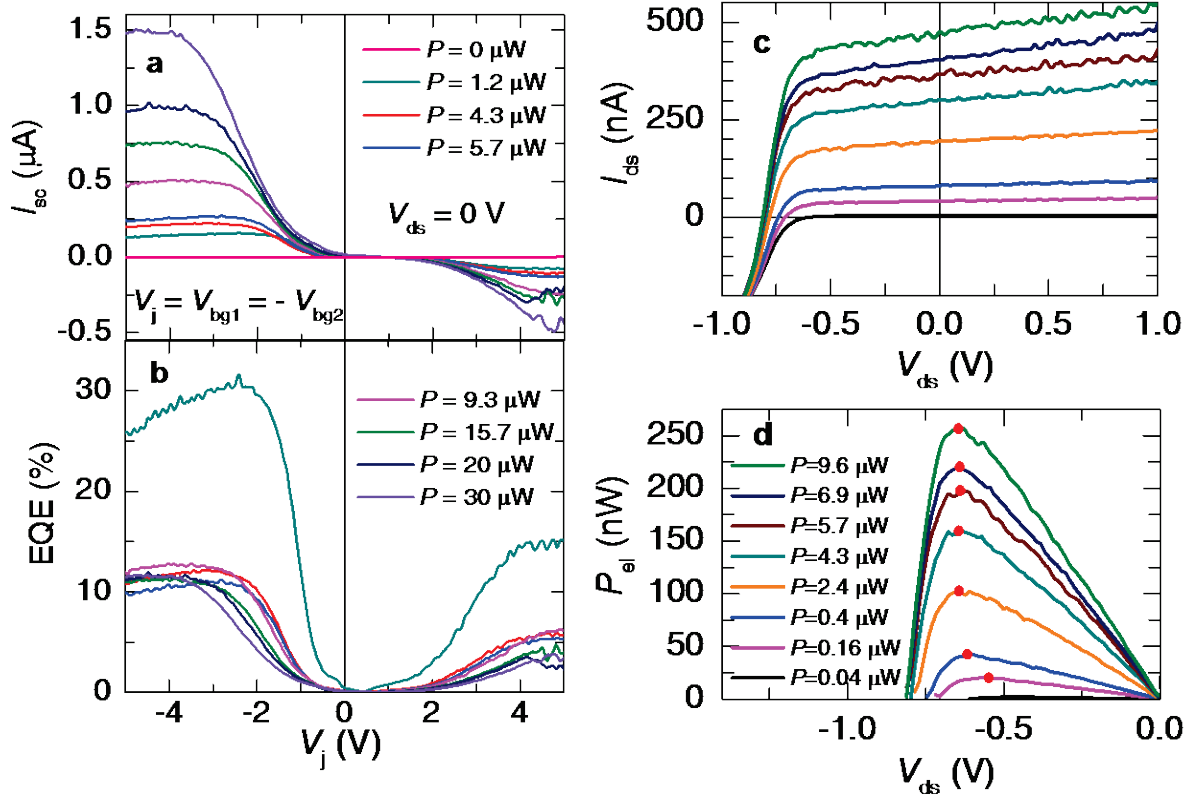

**Figure S4. a** Photo generated short circuit current $I_{sc}$ (or $I_{ph}(V_{ds} = 0\ V)$) for sample #1 as a function of the gate voltage(s) $V_j = V_{bg1} = - V_{bg2}$ and for several values of the illumination power $P$. As seen, under $P = 30$ µW and $V_j = -5$ V one extracts a sizeable photovoltaic current of 1.5 µA, which, when renormalized by the cross sectional area of the sample, translates into a remarkably high current density $j_{sc} \sim 3 \times 10^3$ A/cm$^2$. **b** External quantum efficiencies from the curves in **a**. Notice how it achieves values approaching ~ 30% at the lowest $P$ values. **c** $I_{ds}$ as a function of $V_{ds}$ for sample #2 under $V_j = 5$ V (PN configuration) showing diode-like, or rectification response and for several values of the illumination power $P$. Short circuit currents and open circuit voltages $V_{oc} = V_{ds}(I_{ds} = 0\ A)$ can be directly extracted from the figure. **d** Photo-generated electrical power $P_{el} = I_{ds} \times V_{ds}$ from the second quadrant (i.e. $I_{ds} > 0$ and $V_{ds} < 0$) in **c**. In **c** and in **d** the applied laser power $P$ is indicated by the same line color scheme. Red dots indicate the maximum extracted photovoltaic power levels $P_{el}^{max}$.

Supplementary Figure S5a displays the $I_{sc}$ values as extracted from Fig. S4a under $V_j = -5$ V, while Fig. S5b shows the open-circuit voltages $V_{oc}$ extracted from Fig. S4c, both quantities as functions of the power density $p$. In both graphs, we included also data from the second sample (open circles) measured under lower power levels and when the back-gates were excited under $V_j = +5$ V. If one renormalized these $I_{sc}$ by the channel width (~5 µm) and the crystal thickness (~ 7 nm), to obtain the short-circuit current density $j_{sc}$, one would obtain extremely large values ranging from 1A/cm$^2$ to values in excess of 1 kA/cm$^2$. As for any solar cell, the $j_{sc}$ calculated in this way would indeed represent the *photo-generated* electrical current flowing from the PN-junction and subsequently collected at the electrical contacts. However, and in contrast with our lateral geometry, conventional solar cells are vertical heterostructures for which the incident light vertically illuminates the entire area of the cell. The total short circuit current can be estimated through a simple multiplication of $j_{sc}$ by the area of the cell. In contrast, for our lateral architecture the current flows perpendicularly to the flux of the incident photons, while the area of the junction cannot be easily re-scaled to arbitrarily larger values. Hence, our short circuit current densities have a distinct physical meaning with respect to the conventional definition used in photovoltaics. Nevertheless, they still result from the density of photo-generated electron-hole pairs. Theoretical work is required to understand the significance of these pronounced short-circuit current densities

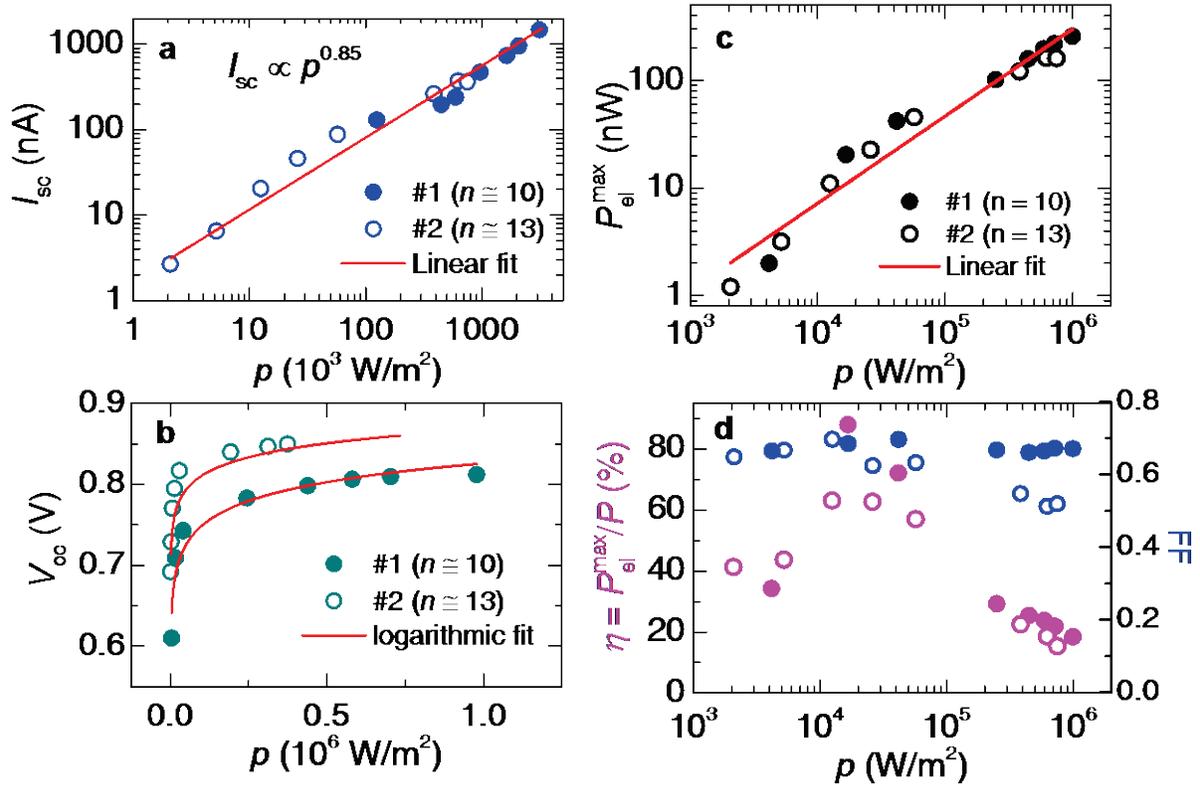

**Figure S5. a** Log-log plot of the photovoltaic or short-circuit current $I_{sc}$ as a function of the laser power density $p$ for $V_j = -5$ V, from the data in Figs. S4a (solid blue circles) and S4c (open circles). Red line is a linear fit yielding $j_{sc} \propto p^{0.85}$. **b** Open-circuit voltage $V_{oc}$ as extracted, for example, from the curves in Fig. S4c as a function of $p$ (solid circles). The same graph contains data from a second sample (open circles). Red lines are logarithmic fits. **c** Maximum extracted photovoltaic output power or $P_{el}^{max}$ from the curves in Fig. S4d as a function of the applied power density $p$. $P_{el}^{max}$ values extracted from samples #1 and #2 are indicated by solid and open circles, respectively. Red line is a linear fit yielding $P_{el}^{max} \propto p^{0.8}$. **d** Photovoltaic efficiency $\eta$ for both samples (magenta markers) and photovoltaic fill-factor FF (blue markers) as functions of $p$. Solid and open circles depict results from samples #1 and #2, respectively.

In Figs. S5a and S5b red lines are a linear (yielding $I_{sc} \propto p^{0.85}$) and logarithmic fits, respectively. The open circuit voltage is expected to follow the equation $V_{oc} = (fk_BT/e)(I_L/I_0+1)$, where $I_L$ is the current generated by light at a given temperature $T$. Notice the large saturating values of $V_{oc} \cong 0.8$ V (sample #1) and 0.85 V (sample #2). Figure S5c displays $P_{el}^{max}$ as a function of $p$ (as extracted from Fig. S4d, where it is indicated by red dots); red line is a linear fit yielding $P_{el}^{max} \propto p^{0.8}$ thus indicating that the photo current is not due to the photo-thermo-electric-effect. Again open markers depict data from sample #2. Finally, Fig. S5d displays the extracted efficiency $\eta = P_{el}^{max}/P$ and the photovoltaic fill factors or FF = $P_{el}^{max} / (I_{sc} \times V_{oc})$, yielding $\eta$ values approaching 40 % at the lowest $p$'s, which peaks to values between 60 % and 80 % as $p$ increases. At the highest illumination power densities, $\eta$ decreases to values ranging from 10 and 20 %. At the moment we do not have a clear physical explanation for the peak in $\eta$ around $p \cong 2 \times 10^4$ W/m². However, it is reproducible and observed in 4 samples characterized in this way. Here, in order to evaluate $\eta$ we normalized the laser power by the laser spot size (3.5 μm in diameter) and multiplied it by the *active area* of the depletion junction placed under illumination; namely its width of 400 nm multiplied by the laser spot diameter. FF on the other hand, remains nearly constant at ~ 0.7, although in sample # 2 it decreases to 0.5 at higher power levels. In contrast to conventional solar cells where the *same* cross-sectional area is used to calculate both the current and the illumination power densities, for our lateral PN-junctions these cross-sectional areas are distinct.

Nevertheless, and as discussed in the main text, we emphasize that the use of the length of the illuminated channel for the calculation of the photovoltaic efficiency, leads to inconsistent and wildly varying values.

## 6. Numerical simulations: role of surface plasmon polaritons at the electrical contacts

To test some of the assumptions in our evaluation of their photovoltaic conversion efficiency, we have carried out a comprehensive numerical analysis of the electromagnetic transmission properties of the experimental devices. The numerical tool of choice was Comsol Multiphysics, a commercially available solver of Maxwell Equations that implements the finite element method. The simulations were performed in a 2D environment, in which translational invariance of the electromagnetic fields along the axis of the gap between gates was imposed. This results in the decoupling of *p*- and *s*-polarizations, which can be treated independently. Perfect Matching layers where used at the edge of the simulation domain, and a mesh size below 1 nm was used to account for the smallest length scales in the system.

In Figure S6, we present the results obtained for the geometry considered in Figure 3 of the main text (see left inset). The right (left) panel corresponds to p-polarized (s-polarized) light. The gap width (height) between top and bottom gates was set to 450 (110) and 240 (35) nm, respectively. The Au permittivity was taken from a multi-Lorentzian fitting of experimental data.[2] The *h*-BN and MoSe$_2$ layer thicknesses were set to 25 and 10 nm, with dielectric constants equal to 4.5 and 13.0, respectively. The spectral transmission was normalized to the electromagnetic power impinging in the slit entrance width (450 nm), and was evaluated at the different material interfaces present in the structure (see right inset).

Figure S6 shows that the normalized-to-area transmission is ~1 for λ>500 nm (energies below 2.5 eV) for both polarizations. Importantly, the transmission integrated to the whole spectral window and evaluated at the Air-MoSe$_2$ interface,

$$T = \frac{1}{700 \text{ nm}} \int_{200 \text{ nm}}^{900 \text{ nm}} T(\lambda) d\lambda,$$

is equal to 0.70 and 0.74 for *p*-polarized and *s*-polarized light, respectively. This indicates that the structure presents a rather small sensitivity to the incident polarization.

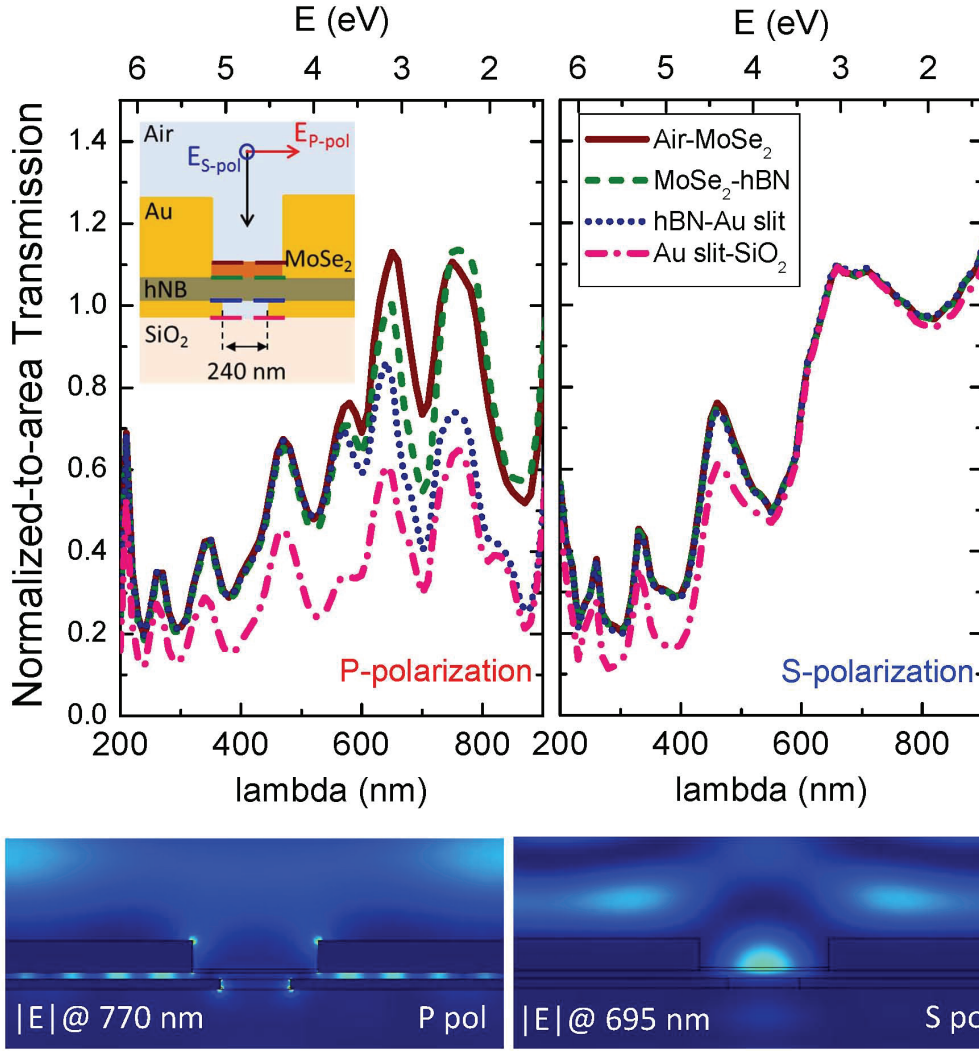

**Figure S6**. Normalized-to-area transmission spectra evaluated for the two possible polarizations of the incident light (*p* and *s*). The transmission is normalized with respect to the light impinging into the gap region between the back-gates. The bottom panels show the electric field amplitudes calculated at the transmission maxima in each case.

Note that the spectral transmission maxima in Figure S6 may have a different physical origin for each polarization. In both cases, the excitation of waveguide modes at the gap between the top Au gates enhances the transmission of the structure. In addition, for *p*-polarized illumination, surface plasmon polaritons (SPPs) can also be excited at the Au interfaces.[3]

The role played by SPPs is investigated in the bottom panels of Figure S6, which renders the electric field amplitude across the experimental geometry, and evaluated at the incident wavelength yielding the highest transmission in the top spectra. As shown in the left panel, SPPs can be detrimental for the photovoltaic conversion efficiency. The electric field map evidences the propagation of SPPs at the *h*-BN layer. These are excited at the gap between the top gates and lead to the enhancement of the electric field amplitude in this layer. However, they also carry electromagnetic power away from the $MoSe_2$ active region. On the contrary, for the case of s-polarization, the transmissivity of the structure is only governed by top gap wave-guide modes, which although weaker, are localized at the active region (see bottom right panel of Figure S6).